\documentclass[12pt,a4paper]{article}

\usepackage{amssymb}
\usepackage{setspace}
\usepackage{indentfirst}

\usepackage{graphicx}

\usepackage{url}
\begin{document}

\begin{titlepage}
\begin{center}

\vspace*{25mm}

\begin{spacing}{1.7}
{\LARGE\bf
 Hubble tension may indicate time-dependent dark matter comoving energy density}
\end{spacing}

\vspace*{25mm}

{\large
Noriaki Kitazawa
}
\vspace{10mm}

Department of Physics, Tokyo Metropolitan University,\\
Hachioji, Tokyo 192-0397, Japan\\
e-mail: noriaki.kitazawa@tmu.ac.jp

\vspace*{25mm}

\begin{abstract}
The values of Hubble constant $H_0$
 by direct measurements with standard distance ladder
 are typically larger than those obtained from
 the observation of cosmic microwave background and the galaxy survey
 with inverse distance ladder.
On the other hand,
 although the errors are still large,
 various determinations of the value of matter density parameter $\Omega_m$
 are consistent with each other.
Therefore,
 it is possible that
 the difference in Hubble constant is translated to the difference
 of physical matter energy density $\omega_m \equiv \Omega_m h^2$,
 where $h \equiv H_0 / (100 \, \mbox{\rm Km/s/Mpc})$.
In this article 
 it is examined
 the possibility of an increase of the physical dark matter energy density
 (comoving energy density without the effect of expansion of the universe)
 by a fast transition at a certain value of redshift
 as a possible resolution of the Hubble tension.
A phenomenological fluid model of the dark sector,
 which is the modification of a so-called unified dark matter model,
 is introduced to concretely realize such a fast transition
 in the physical dark matter energy density.
\end{abstract}

\end{center}
\end{titlepage}

\doublespacing

\section{Introduction}
\label{sec:introduction}

Hubble tension is one of the important problems
 which may give some hints to understand the universe beyond the $\Lambda$CDM model.
The typical values of Hubble parameter in tension are between
 that from direct measurement with standard distance ladder
 $H_0 = 73.04 \pm 1.04 \,\, \rm{Km/s/Mpc}$ \cite{Riess:2021jrx}
 and that from the observation of cosmic microwave background (CMB)
 with inverse distance ladder $H_0 = 67.4 \pm 0.5 \,\, \rm{Km/s/Mpc}$
 \cite{Planck:2018vyg}.
The difference of these two values
 in almost the same good precision represents this problem clearly,
 and much effort has been devoted to solve the problem
 (for review see
 \cite{Knox:2019rjx,DiValentino:2021izs,Schoneberg:2021qvd,Vagnozzi:2023nrq}).
Though some unknown systematic errors in observations
 could resolve the problem \cite{Freedman:2023jcz},
 it is important that many direct observations with different methods
 tend to give larger values of $H_0$ than those from CMB observations.
In this work we consider that
 this problem suggests some new physics beyond the $\Lambda$CDM model.

There are two typical new physics which have been extensively considered:
 one is the early-time physics
 which changes the value of the sound horizon at recombination
 (see \cite{Poulin:2018cxd,Niedermann:2020dwg,Escudero:2019gvw,Brinckmann:2022ajr}
 for typical examples),
 and the other is the physics which changes the way of expansion of the universe
 at late time
 (see \cite{Benevento:2020fev,
 DiValentino:2019jae,Li:2020ybr,Yang:2021eud,Zhou:2021xov}
  for typical examples).
The first one is strongly constrained by CMB observations \cite{Vagnozzi:2021gjh}
 and the second one, especially with the time-dependent dark energy density,
 does not produce enough large effect \cite{Kitazawa:2023peg}.
Both of these two new physics may simultaneously exist \cite{Vagnozzi:2023nrq},
 but unfortunately we have not yet found any convincing solution\footnote{
 A general model independent analysis of these two typical new physics
 is given in \cite{Vagnozzi:2019ezj}.
 The third possibility of local new physics has been investigated: see
 \cite{Giani:2023aor} for example.
 }.

In the observations to obtain the value of Hubble constant
 the matter density parameter $\Omega_m = \rho_m/\rho_{\rm cr}$ is also obtained,
 where $\rho_m$ is the matter (dark matter and conventional matter) energy density
 at present and $\rho_{\rm cr} \equiv 3H_0^2/8\pi G_N$
 is the critical energy density at present.
Since the value of $\Omega_m$
 determines the expansion of the universe in matter-dominated era,
 the resultant observational values are consistent with each other,
 though the errors are still large.
Under the assumption of a common true value of $\Omega_m$
 the values of physical matter energy density parameter
 $\omega_m \equiv \Omega_m h^2$,
 which is independent from the value of Hubble constant,
 should give larger/smaller values in the observations
 which result larger/smaller values of the Hubble constant.
This translate the problem of Hubble tension to the different problem:
 the physical matter energy density
 should be larger at late time than that at early time \cite{Blanchard:2022xkk}.

In the next section
 we carefully investigate the various observational results of
 $\Omega_m$ and $\omega_m = \Omega_m h^2$.
We see that
 there is a tendency of larger $\omega_m$ at late time than that at early time
 with a maximum difference at $2.5 \, \sigma$.
The idea to solve the problem of Hubble tension
 by introducing time-dependent physical matter energy density
 (comoving energy density without the effect of expansion of the universe),
 namely the time-dependence of the parameter $\omega_m$, is introduced
 as a physics beyond the $\Lambda$CDM model.
In section \ref{sec:UDE}
 we introduce a phenomenological fluid model for the dark sector
 (dark matter and dark energy),
 which concretely represents a fast increase of physical dark matter energy density
 at a certain redshift and gives a possible solution of the problem
 at least in the level of background evolution.
In the last section we discuss necessary future works and conclude.

\section{Hubble tension and matter energy density}
\label{sec:DM-density}

We first carefully investigate
 the values of $\Omega_m$ and $\omega_m = \Omega_m h^2$
 by five typical observations:
 three are CMB observations
 \cite{WMAP:2012nax,Planck:2018vyg,ACT:2020gnv},
 three are galaxy survey catalogues
 \cite{Philcox:2021kcw,DES:2021wwk,Kilo-DegreeSurvey:2023gfr},
 and two are type Ia supernova catalogues with local distance ladders
 \cite{Pan-STARRS1:2017jku,Riess:2020fzl,Brout:2022vxf,Riess:2021jrx}.

\begin{figure}[t]
\centering
\includegraphics[width=75mm]{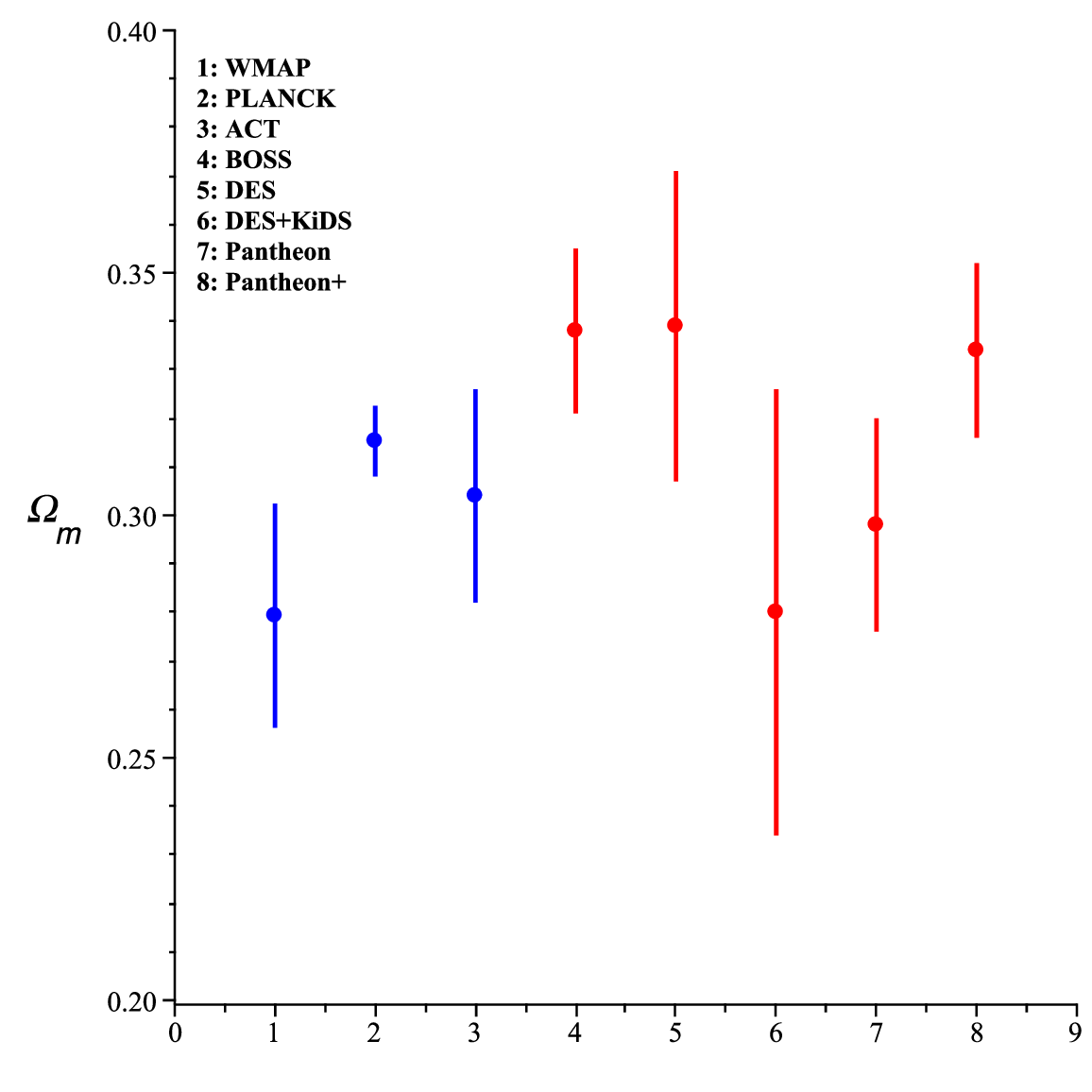}
\quad
\includegraphics[width=75mm]{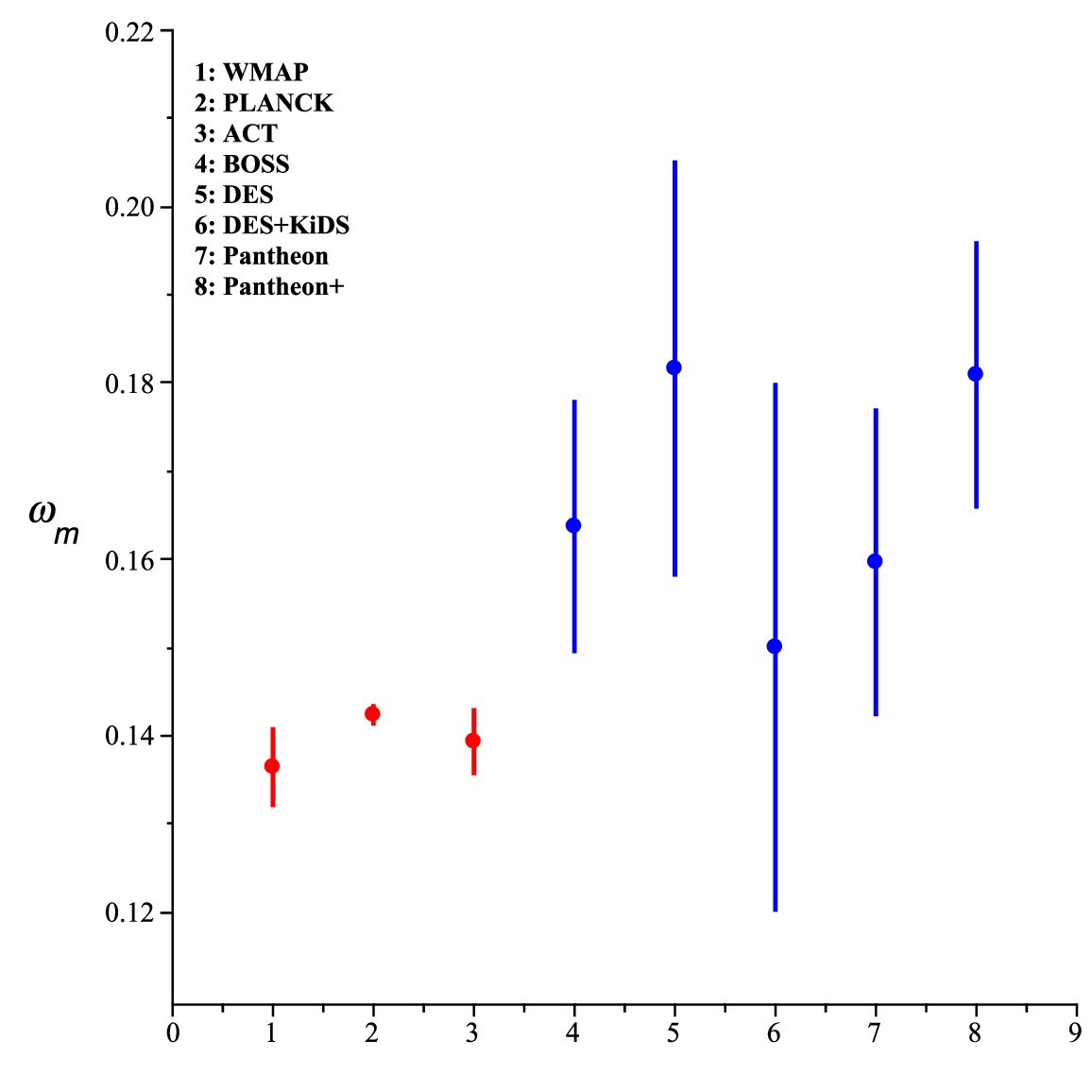}
\caption{
The observed values of $\Omega_m$ (left panel)
 and $\omega_m \equiv \Omega_m h^2$ (right panel) by various observations.
The numbers in the horizontal axes correspond to the observations, namely,
 1: WMAP 9yr, 2: PLANCK 2018, 3: ACT DR4, 4: BOSS DR12 full-shape,
 5: DES Y3, 6: DES + KiDS, 7: Pantheon, and 8: Pantheon+.
The values in red are the results of less model dependent determinations,
 and those in blue are obtained by
 some assumptions mainly assuming $\Lambda$CDM model.
}
\label{fig:matter-density}
\end{figure}

The observation of CMB perturbations gives very precise value of matter energy density
 $\omega_m = \omega_c + \omega_b$,
 where $\omega_c$ and $\omega_b$ are physical cold-dark matter density
 and baryon density, respectively.
Then, the values of $\Omega_m$
 are derived with some assumptions mainly assuming $\Lambda$CDM model.
The results by three observations are
\begin{eqnarray}
 \omega_m &=& 0.1364 \pm 0.0045, \quad  \Omega_m = 0.280 \pm 0.023,
 \qquad \mbox{\rm by WMAP \cite{WMAP:2012nax}},\\
 \omega_m &=& 0.142 \pm 0.0010, \quad  \Omega_m = 0.315 \pm 0.007,
 \qquad \mbox{\rm by PLANCK \cite{Planck:2018vyg}},\\
 \omega_m &=& 0.139 \pm 0.0038, \quad  \Omega_m = 0.304 \pm 0.022,
 \qquad \mbox{\rm by ACT \cite{ACT:2020gnv}},
\end{eqnarray}
 which appear in fig.\ref{fig:matter-density}
 corresponding to 1,2, and 3 in horizontal axes, respectively.

The observation of the real space distribution of galaxies
 and also the observations of the galaxy clustering and weak lensing effects
 can give constraint to the way of expanding the universe.
These observations give value of $\Omega_m$ and $\omega_m$ as
\begin{eqnarray}
 \Omega_m &=& 0.338^{+0.016}_{-0.017}, \quad  \omega_m = 0.164^{+0.013}_{-0.014},
 \qquad \mbox{\rm by BOSS full-shape \cite{Philcox:2021kcw}},\\
 \Omega_m &=& 0.339^{+0.032}_{-0.031}, \quad  \omega_m = 0.182^{+0.024}_{-0.023},
 \qquad \mbox{\rm by DES Y3 \cite{DES:2021wwk}},\\
 \Omega_m &=& 0.280^{+0.037}_{-0.046}, \quad  \omega_m = 0.150^{+0.025}_{-0.030},
 \qquad \mbox{\rm by DES+KiDS \cite{Kilo-DegreeSurvey:2023gfr}},
\end{eqnarray}
 where for DES Y3 and DES+KiDS
 the value of Hubble constant
 $H_0 = 73.2 \pm 1.3\,\, \rm{Km/s/Mpc}$ \cite{Riess:2020fzl}
 is used following that DES Y3 uses this value in \cite{DES:2021wwk}.
Note that
 the BOSS DR12 full-shape directly gives the value $h=0.696^{+0.011}_{-0.013}$,
 which is smaller than those from Pantheon and Pantheon+ supernova catalogues
 \cite{Riess:2020fzl,Riess:2021jrx},
 even though they are all late-time measurements.
In fig.\ref{fig:matter-density}
 these results appear corresponding to 4,5, and 6 in horizontal axes, respectively.

The distance-redshift relation by the observations of type Ia supernovae
 directly gives the values of $\Omega_m$,
 and it also gives the value of Hubble constant with standard distance ladder.
Two typical observations give the result as
\begin{eqnarray}
 \Omega_m &=& 0.298 \pm 0.022, \quad  \omega_m = 0.159 \pm 0.018,
 \qquad \mbox{\rm by Pantheon \cite{Pan-STARRS1:2017jku}},\\
 \Omega_m &=& 0.334 \pm 0.018, \quad  \omega_m = 0.180 \pm 0.015,
 \qquad \mbox{\rm by Pantheon+ \cite{Brout:2022vxf}},
\end{eqnarray}
 where for the value of Hubble constant,
 Pantheon uses the value of \cite{Riess:2020fzl}
 and Pantheon+ uses the value of \cite{Riess:2021jrx}.
In fig.\ref{fig:matter-density}
 these results appear corresponding to 7 and 8 in horizontal axes,
 respectively\footnote{
 For direct determinations of $\Omega_m$ and $H_0$ at higher values of redshift,
 see \cite{Colgain:2022nlb,Colgain:2022rxy,Lenart:2022nip,Dainotti:2022bzg},
 for example.
 }.

In the left panel of fig.\ref{fig:matter-density}
 we see that typical present available results of $\Omega_m$
 are consistent with each other, though errors are still large
 \cite{Sakr:2023hrl}.
This indicates that
 the observation, which results larger value of the Hubble constant,
 gives also larger value of physical $\omega_m \equiv \Omega_m h^2$,
 as it can be seen in the right panel of fig.\ref{fig:matter-density}.
Though BOSS DR12 full-shape gives rather small Hubble constant,
 the corresponding value of physical matter energy density is rather large,
 because their value of $\Omega_m$ is relatively large.
There is a clear tendency that
 the physical matter energy density is smaller in CMB observations
 than that in late-time observations.
If we compare the values from Planck 2018 with Pantheon+,
 the difference in physical matter energy density is about $2.5 \, \sigma$.
This could indicate time-dependent physical dark matter energy density
 (comoving energy density without the effect of expansion of the universe),
 namely the value is smaller at larger redshifts.
In the following we neglect the baryon contribution to the matter density,
 since the contribution is small and the possible unknown physics
 should be in the dark sector.

\begin{figure}[t]
\centering
\includegraphics[width=75mm]{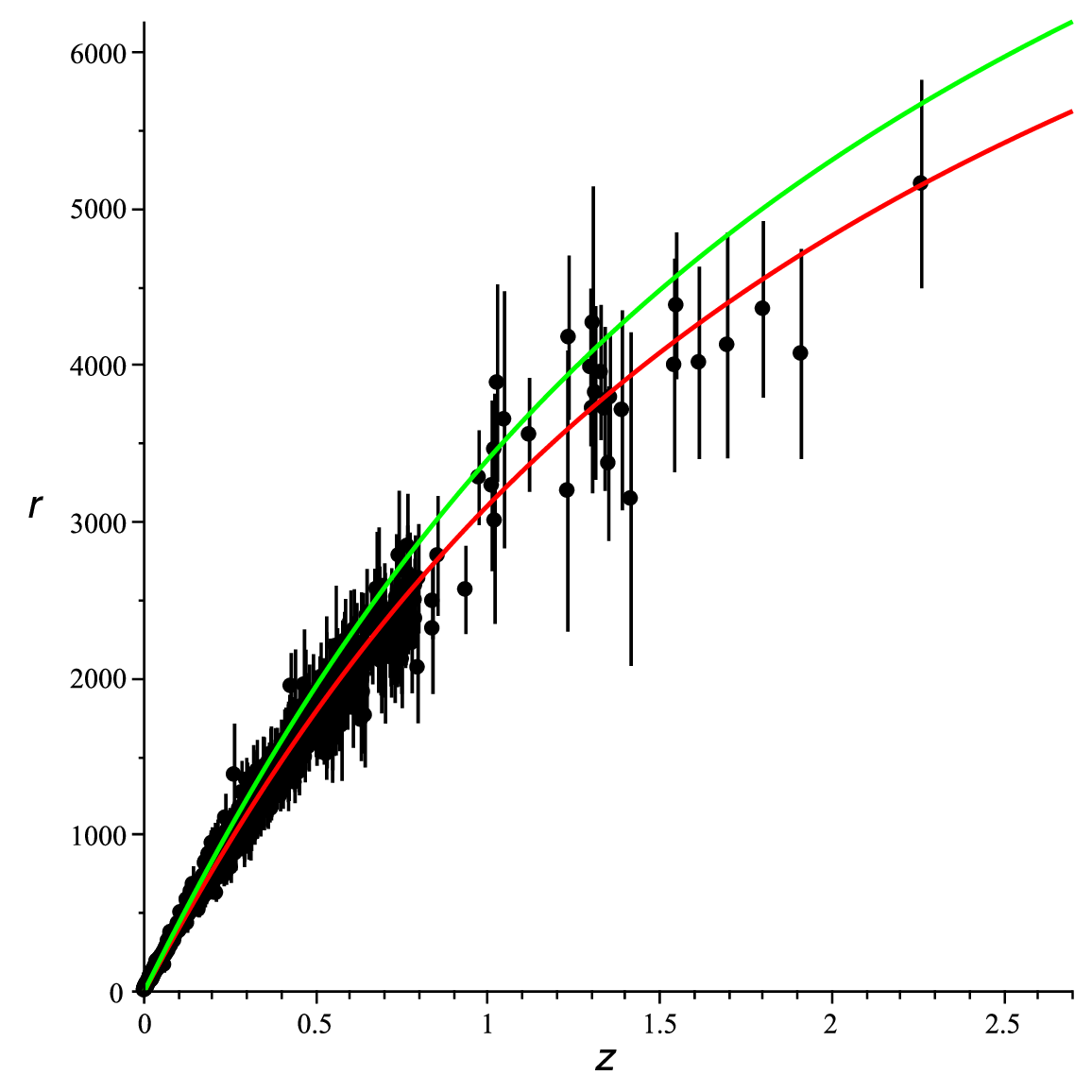}
\quad
\includegraphics[width=75mm]{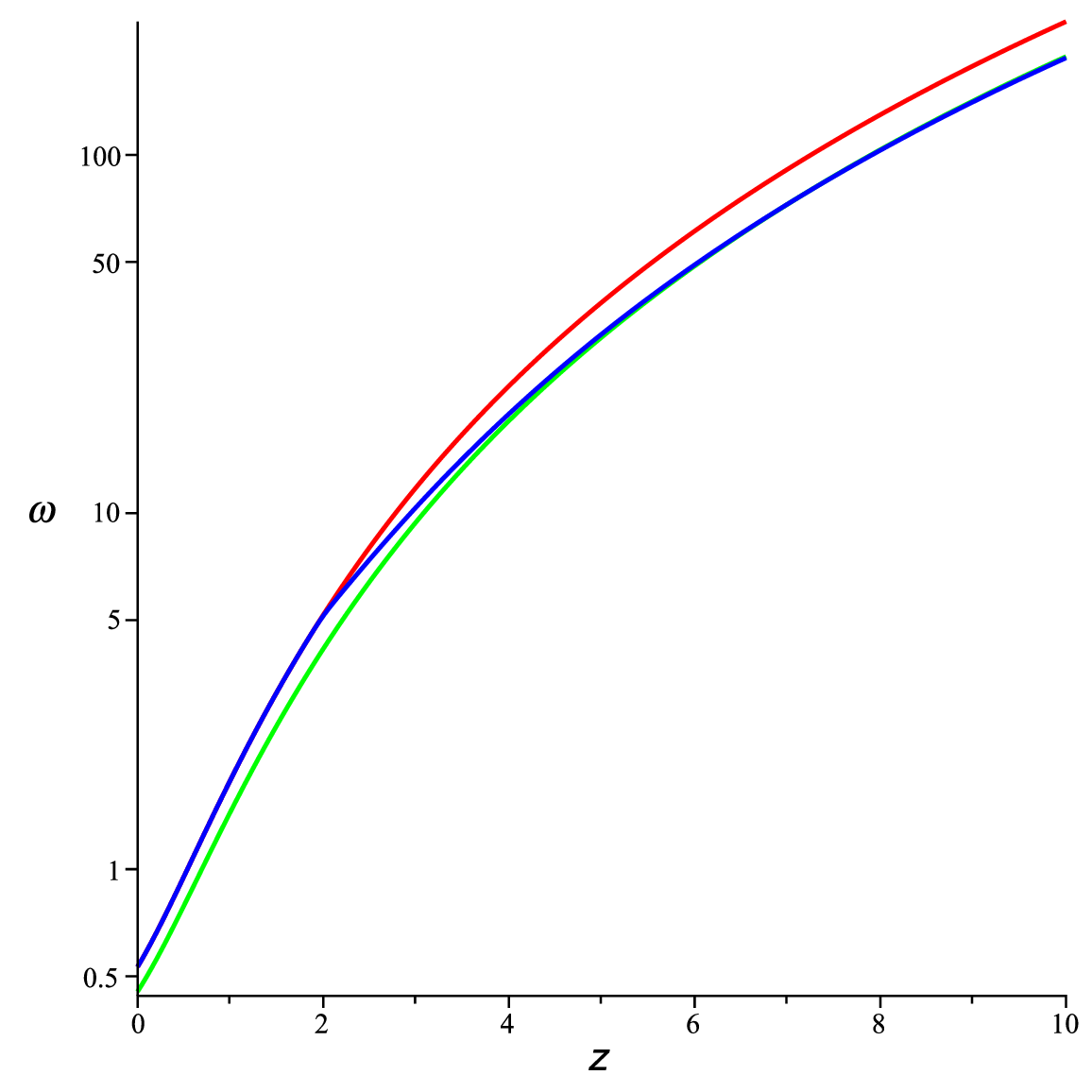}
\caption{
Left panel:
Distance-redshift relation.
Black dots with error bars indicate Pantheon+ supernova data.
The red line indicates the prediction of the SH0ES $\Lambda$CDM model
 and green line indicates the prediction of the Planck $\Lambda$CDM model.
Right-panel:
Evolution of physical total energy density,
 including the effect of the expansion of the universe,
 $\omega(z) = (H(z) / (100 \, \mbox{\rm Km/s/Mpc}))^2$,
 where colors red and green denote the same in the left panel.
The blue line indicates the prediction of a model
 with time-dependent physical dark matter energy density.
}
\label{fig:distance-redshift}
\end{figure}

Next, we summarize the problem of Hubble tension in a simple way.
In the following
 we consider only two typical values of Hubble constant
 $H_0^{\rm late} \equiv 73.04 \pm 1.04 \,\, \rm{Km/s/Mpc}$ by SH0ES
 \cite{Riess:2021jrx}
 and  $H_0^{\rm early} \equiv 67.4 \pm 0.5 \,\, \rm{Km/s/Mpc}$ by Planck
 \cite{Planck:2018vyg}.
The solution of the problem
 is not the reconciliation of these two extreme values,
 but it is to understand the fact that many direct observations with different methods
 tend to give larger values than those from CMB observations.
In this work, however,
 we consider only these two extreme values to clearly show our point.
We consider the following two $\Lambda$CDM models of Hubble parameter evolution
\begin{equation}
 H(z) = H_0 \sqrt{\Omega_\Lambda + \Omega_m(1+z)^3}
\end{equation}
 by taking different sets of values of parameters:
 for ``SH0ES $\Lambda$CDM model''
\begin{equation}
 H_0 = H_0^{\rm late},
 \quad \Omega_m = \Omega_m^{\rm late} = 0.334,
 \quad \Omega_\Lambda = \Omega_\Lambda^{\rm late} = 1 - \Omega_m^{\rm late}
\end{equation}
 and for ``Planck $\Lambda$CDM model''
\begin{equation}
 H_0 = H_0^{\rm early},
 \quad \Omega_m = \Omega_m^{\rm early} = 0.315,
 \quad \Omega_\Lambda = \Omega_\Lambda^{\rm early} = 1 - \Omega_m^{\rm early},
\end{equation}
 where we are assuming flat space-time.
The distance-redshift relation from type Ia supernovae is described as
\begin{equation}
 r(z) = \frac{1}{1+z} \, 10^{(m(z)-(25+M))/5} \,\, \mbox[\rm Mpc],
\end{equation}
 where $m(z)$ is the apparent magnitude of a supernova at $z$
 and we take $M = -19.2$ following the SH0ES distance calibration
 with standard distance ladder \cite{Riess:2021jrx}.
Theoretically the distance-redshift relation is simply given by
\begin{equation}
 r(z) \equiv \int_0^z \frac{dz'}{H(z')}.
\label{distance-theory}
\end{equation}
In this work the distance $r(z)$ is the simple light propagation distance
 related to the luminosity distance $d_L(z) = (1+z) r(z)$.
The left panel of fig.\ref{fig:distance-redshift}
 shows distance-redshift relation by Pantheon+ catalogue (black dots with error bars)
 with the prediction by the SH0ES $\Lambda$CDM model in red line
 and also the prediction by the Planck $\Lambda$CDM model in green line.
It clearly shows that
 the SH0ES $\Lambda$CDM model
 fits well the Pantheon+ catalogue with SH0ES distance calibration.
The Planck $\Lambda$CDM model,
 which is obtained by the extrapolation from the early-time physics observation,
 does not fit the late-time Pantheon+ catalogue with SH0ES distance calibration.
The main difference of these two models are the values of Hubble parameter,
 namely the normalization of eq.(\ref{distance-theory}),
 and the difference of matter density parameter gives small effects.
This is a simple summary of the problem of Hubble tension.

Now we introduce the idea to solve this problem
 by introducing time-dependent physical matter energy density.
We need to accept the SH0ES $\Lambda$CDM model at least $z \lesssim 2$,
 but at higher redshifts $z > z_t$,
 where $z_t > 2$ is some transition point,
 the Planck $\Lambda$CDM model should be realized
 to be consistent with early-time physics observation.
From the first argument in this section
 we assume that the value of $\Omega_m$ is unique and therefore
\begin{equation}
 \Omega_m
  = \omega_m^{\rm late}/(h^{\rm late})^2
  = \omega_m^{\rm early}/(h^{\rm early})^2,
\end{equation}
 where $h^{\rm late} \equiv H_0^{\rm late} / (100 \, \mbox{\rm Km/s/Mpc})$
 and $h^{\rm early} \equiv H_0^{\rm early} / (100 \, \mbox{\rm Km/s/Mpc})$.
We also assume that
 the value of physical matter energy density changes
 from $\omega_m^{\rm late}$ to $\omega_m^{\rm early}$ at $z_t$
 for larger redshifts.
In this case at large redshifts $z \gg z_t$
\begin{eqnarray}
 H(z)
 &=& H_0^{\rm late}
   \sqrt{\Omega_\Lambda^{\rm late} + \Omega_m^{\rm late} (1+z)^3}
 = H_0^{\rm late}
   \sqrt{\Omega_\Lambda^{\rm late} + (h^{\rm late})^{-2} \omega_m^{\rm late} (1+z)^3}
\nonumber\\
 &\rightarrow&
   H_0^{\rm late}
   \sqrt{(h^{\rm late})^{-2} \omega_m^{\rm early} (1+z)^3}
 = H_0^{\rm late}
   \sqrt{\left(\frac{h^{\rm early}}{h^{\rm late}}\right)^2 \Omega_m (1+z)^3}
\nonumber\\
 &=& H_0^{\rm early}
   \sqrt{\Omega_m (1+z)^3}.
\end{eqnarray}
In this way the Planck $\Lambda$CDM model is realized at larger redshifts
 where we can safely neglect the contribution of the dark energy.
The right panel of fig.\ref{fig:distance-redshift} shows
 the redshift-dependence of the total physical energy density
 $\omega(z) = (H(z) / (100 \, \mbox{\rm Km/s/Mpc}))^2$.
Note that it takes the value of $h^2$ at $z=0$.
The red line is that for the SH0ES $\Lambda$CDM model
 and green line is that for the Planck $\Lambda$CDM model.
The blue line is that
 for the model with the time-dependence of physical dark matter energy density,
 which we will introduce in the next section.
The blue line bridges two lines of SH0ES and Planck,
 and this behavior is very different from those of the energy densities
 in the models with non-trivial time-dependence only on the dark energy densities,
 since the dark energy dominates the universe only $z < 0.3$
 (see \cite{Kitazawa:2023peg} for example).

The change of the behavior of $H(z)$
 alters the value of comoving angular diameter distance
 to the last scattering surface, $d_A^*$,
 which is tightly constrained as $d_A^*=r_s^*/\theta_*$,
 where $r_s^*$ is the size of the sound horizon at the last scattering surface
 (redshift $z^*$) and $\theta_*$ is the angular acoustic scale in CMB perturbations.
We may roughly write
\begin{equation}
 d_A^* = \int_0^{z_t} dz \, \frac{1}{H_{\rm SH0ES}(z)}
       + \int_{z_t}^{z^*} dz \, \frac{1}{H_{\rm Planck}(z)},
\end{equation}
 where $H_{\rm SH0ES}(z)$ and $H_{\rm Planck}(z)$ indicate
 Hubble parameters of SH0ES and Planck $\Lambda$CDM models, respectively.
This formula should give the value of $r_s^*/\theta_*$ in case of $z_t=0$.
For an appropriate finite value of $z_t$ the value of $d_A^*$ becomes smaller,
 since the first term of the above equation gives smaller contribution,
 because $H_{\rm SH0ES}(z) > H_{\rm Planck}(z)$
 due to $H_0^{\rm late} > H_0^{\rm early}$.
Therefore,
 some new physics may be required to reduce the value of $r_s^*$.
The possible new physics may be some early-time physics,
 which have been mentioned in the first section,
 or some other late-time physics like a sign switching cosmological constant
 \cite{Akarsu:2019hmw,Akarsu:2021fol,Akarsu:2022typ,Akarsu:2023mfb}.

\section{Applications of a unified dark matter model}
\label{sec:UDE}

We introduce
 a phenomenological fluid model for the dark sector (dark matter and dark energy),
 which provides a fast transition of physical dark matter energy density,
 to realize the proposed idea in the previous section.
The model is based on a unified dark matter model
 \cite{Bertacca:2010mt,Frion:2023xwq}
 which describes both the dark matter and dark energy in one fluid
 by the late-time emergence of the dark energy.
Here, we are going to use the idea inversely
 to emergent the dark matter from the dark energy.

Consider a perfect fluid with energy density $\rho$ and pressure $p$
 in flat Friedmann‐Lema\^itre‐Robertson-Walker metric with scale parameter $a$.
In the following we concentrate on the dark sector only.
The conservation law of this fluid for the dark sector can be described as
\begin{equation}
 a \frac{d\rho}{da} + 3 \rho = -3p.
\end{equation}
The formal solution of this equation with a given $p(a)$ is
\begin{equation}
 \rho(a) = \frac{1}{a^3}
 \left[
  K - 3 \int_0^a d\bar{a} \,\, {\bar{a}}^2 \,\, p({\bar a})
 \right],
\end{equation}
 where $K$ is an integration constant
 which interestingly describes some energy density of non-relativistic matter.
The model is specified by setting a pressure function of $p(a)$ and we set
\begin{equation}
 p(a) = - \Lambda
        - \frac{\rho_\lambda}{2}
          \left[
           1 - \tanh \left\{ \frac{\beta}{3} \left( a^3 - a_t^3 \right) \right\}
          \right].
\label{model-p}
\end{equation}
This is a modification of the model in \cite{Frion:2023xwq}
 by introducing a constant $\Lambda$ and changing the sign
 in front of $\tanh$ function.
The parameters $\beta$ and $a_t$ describe
 the quickness and the scale factor (or the time)
 of a transition in pressure, respectively.
We normalize the scale factor as $a(t_0)=1$,
 and the redshift $z$ is described as $1+z=1/a$.
The solution of the conservation equation can be written as
\begin{equation}
 \rho(a) = \left( \Lambda + \frac{\rho_\lambda}{2} \right)
         + \frac{{\tilde \rho}_m}{a^3}
         - \frac{\rho_\lambda}{2} \frac{1}{a^3} \frac{3}{\beta}
           \ln\cosh \left\{ \frac{\beta}{3} \left( a^3 - a_t^3 \right) \right\},
\label{model-rho}
\end{equation}
 where the integration constant $K$
 is chosen to give a simple term of ${\tilde \rho}_m/a^3$,
 where ${\tilde \rho}_m$ is a constant.
In the following we consider the fast transition $\beta=1000$
 and set the transition point at $z_t=2$, namely $a_t=1/3$, as an example.
The remaining three parameters,
 $\Lambda$, ${\tilde \rho}_m$ and $\rho_\lambda$, can be fixed as follows.

In case of fast transition
 the quantity $\alpha \equiv \frac{\beta}{3} (a^3-a_t^3)$
 is always large in magnitude
 except for the region around the transition point $a = a_t$.
To investigate the energy components before and after the transition
 we consider the asymptotic behavior for large $\vert\alpha\vert$ of
 eqs.(\ref{model-p}) and (\ref{model-rho}).
\begin{eqnarray}
 p &\simeq&
  - \left( \Lambda + \frac{\rho_\lambda}{2} \right)
  + \frac{\rho_\lambda}{2} \, \mbox{sgn}(\alpha),
\\
 \rho &\simeq&
    \left( \Lambda + \frac{\rho_\lambda}{2} \right)
  + \left(
     {\tilde \rho}_m + \frac{\rho_\lambda}{2} \frac{3\ln2}{\beta}
    \right) \frac{1}{a^3}
  - \frac{\rho_\lambda}{2} \left\vert 1 - \frac{a_t^3}{a^3} \right\vert,
\end{eqnarray}
 where $\mbox{sgn}$ is the sign function.
The energy density before the transition ($a < a_t$) is
\begin{equation}
 \rho \simeq \left( \Lambda + \rho_\lambda \right)
  + \left(
     {\tilde \rho}_m + \frac{\rho_\lambda}{2} \frac{3\ln2}{\beta}
     - \frac{\rho_\lambda a_t^3}{2}
    \right) \frac{1}{a^3},
\label{before-trans-rho}
\end{equation}
 and that after the transition ($a > a_t$) is
\begin{equation}
 \rho \simeq \Lambda
  + \left(
     {\tilde \rho}_m + \frac{\rho_\lambda}{2} \frac{3\ln2}{\beta}
     + \frac{\rho_\lambda a_t^3}{2}
    \right) \frac{1}{a^3},
\label{after-trans-rho}
\end{equation}
 which clearly shows an energy transition from dark energy ($\sim a^0$)
 to the non-relativistic matter ($\sim a^{-3}$) with $\rho_\lambda>0$.
Note that
 these equations can be understood as the formula of energy density
 in the limit of instantaneous transition, since they coincide at $a=a_t$.
Numerically,
 we can fix three parameters of $\Lambda$, ${\tilde \rho}_m$ and $\rho_\lambda$
 so that eq.(\ref{after-trans-rho})
 coincides with that of the SH0ES $\Lambda$CDM model,
 and also the second term of eq.(\ref{before-trans-rho})
 coincides with that of the Planck $\Lambda$CDM model.
The blue line in the right panel of fig.\ref{fig:distance-redshift}
 has been obtained in this way,
 and the total energy density, eq.(\ref{model-rho}),
 is always positive at all values of redshift.

\begin{figure}[t]
\centering
\includegraphics[width=75mm]{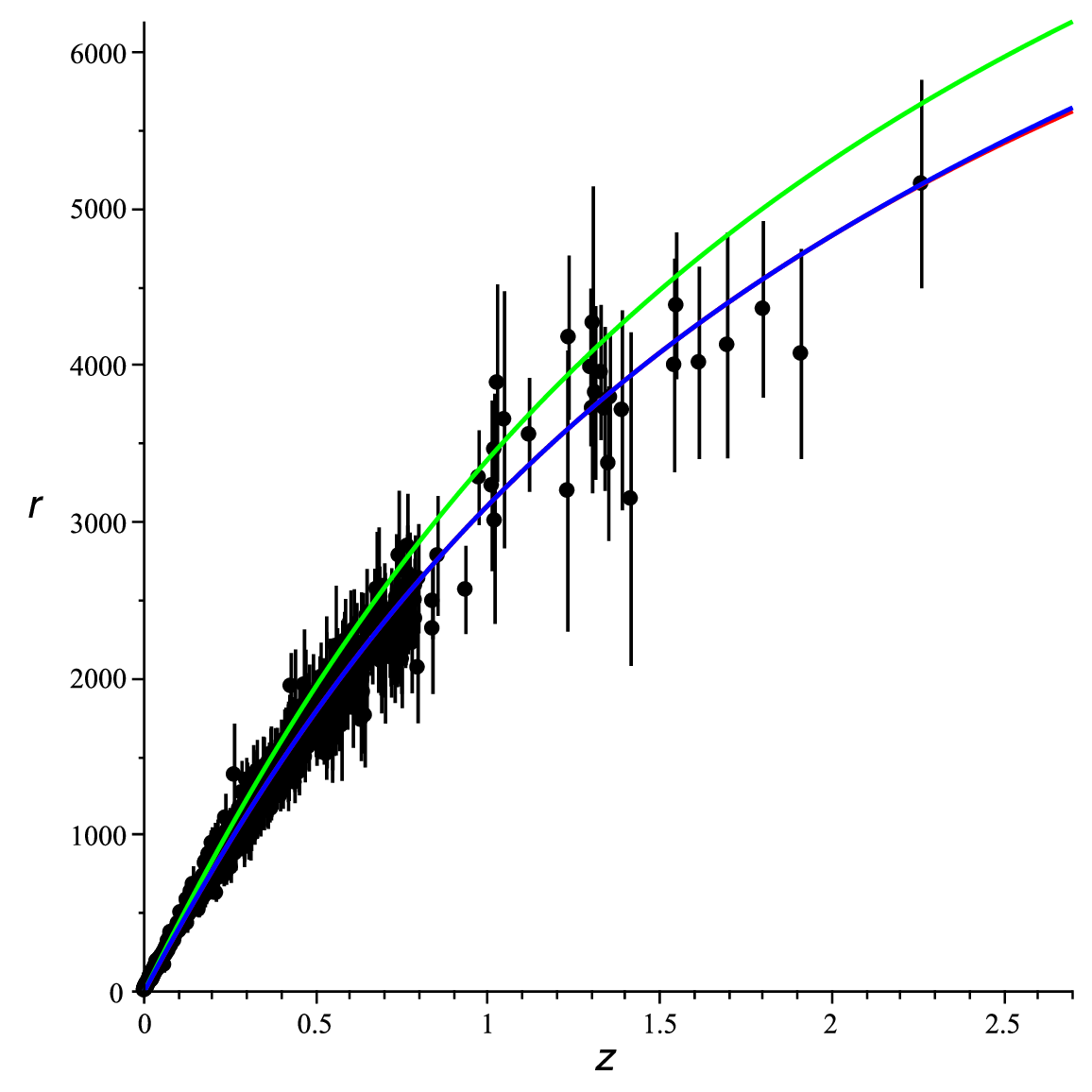}
\quad
\includegraphics[width=75mm]{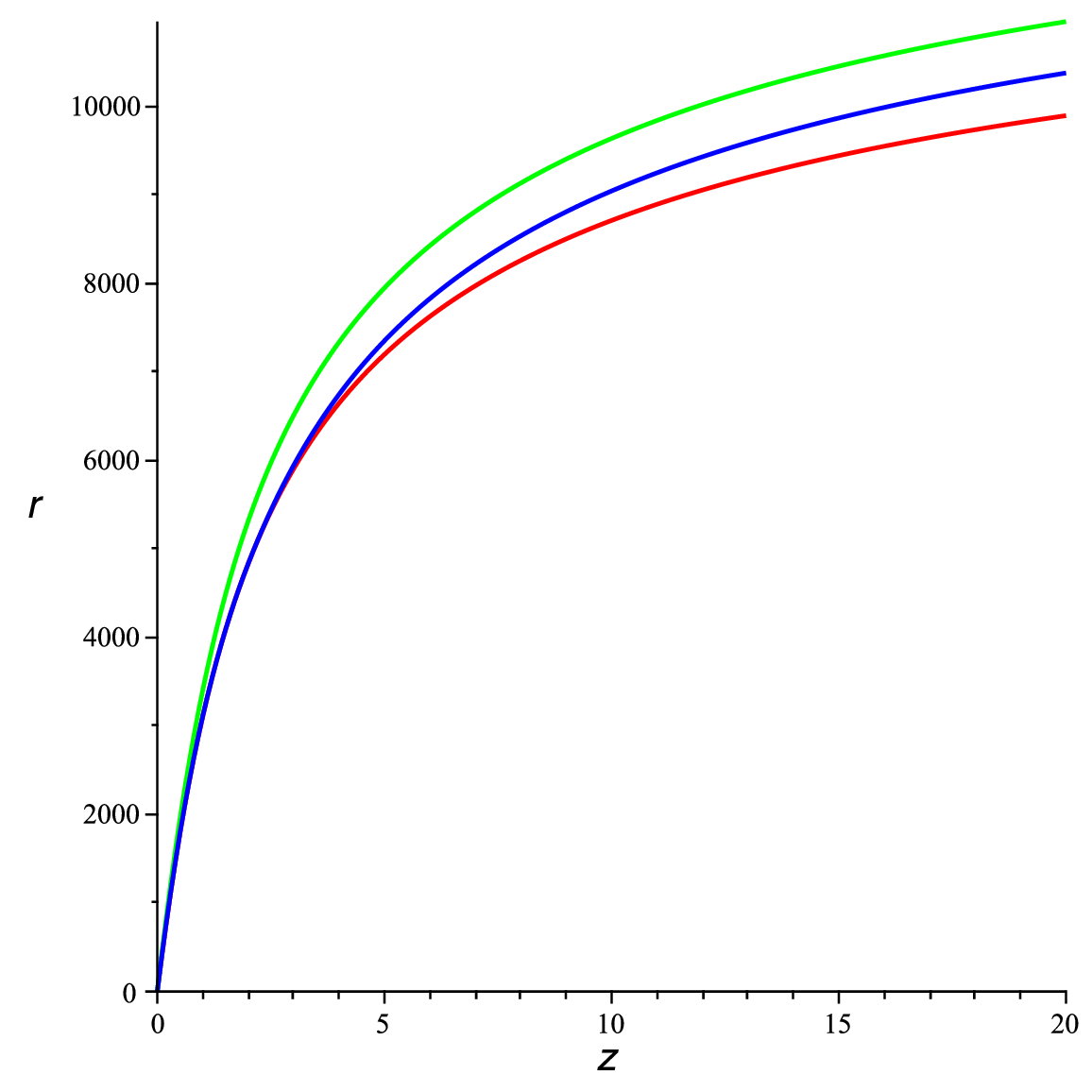}
\caption{
Distance-redshift relation.
Left panel:
Black dots with error bars indicate Pantheon+ supernova data.
The red line indicates the prediction of the SH0ES $\Lambda$CDM model
 and green line indicates the prediction of the Planck $\Lambda$CDM model.
The blue line, which indicates the prediction of our fluid model,
 almost overlaps with the red line.
Right-panel:
Theory predictions for larger redshift values.
The color convention is the same of that in the left panel.
}
\label{fig:distance-redshift-model}
\end{figure}

In fig.\ref{fig:distance-redshift-model}
 we compare the predictions to the distance-redshift relation
 by three models: the SH0ES $\Lambda$CDM model (red line),
 the Planck $\Lambda$CDM model (green line),
 and the fluid model (blue line).
We see in the left panel that
 the fit of the fluid model with Pantheon+ catalogue with SH0ES distance calibration
 is very good.
In the right panel
 we see how the distance-redshift relation of the fluid model
 leaves from that of the SH0ES $\Lambda$CDM model at low redshift values
 to that of the Planck $\Lambda$CDM model at higher redshift values.

This fluid model
 is a concrete realization of the idea which is proposed in the previous section.
At very large redshifts around the period of recombination
 the Planck $\Lambda$CDM model works very well for CMB physics
 (the value of dark energy density is irrelevant at that time).
The universe evolves following the Planck $\Lambda$CDM model
 until at a certain redshift where a fast transition happens.
In the fast transition
 a part of the dark energy density transforms to the dark matter energy density,
 which increases the amount of physical dark matter energy density
 (comoving energy density without the effect of expansion of the universe).
After the transition 
 the universe evolves following the SH0ES $\Lambda$CDM model at low redshifts.

\section{Discussions and conclusions}
\label{sec:conclusions}

Since the fact that
 the various observations give consistent values of matter density parameter
 $\Omega_m$,
 which determines the way of expansion of the universe in matter-dominated era,
 Hubble tension could be translated to the problem of the discrepancy in
 physical matter energy density parameter $\omega_m \equiv \Omega_m h^2$.
There is a tendency that
 the values of $\omega_m$ from late-time observations
 are larger than those from early-time observations.
This may indicate that
 the physical dark matter energy density increases at a certain redshift value
 in the expansion history of the universe.
This is obviously the physics beyond the $\Lambda$CDM model.
It has been shown that
 this possibility can be concretely realized in a fluid model,
 as a fast transformation of the dark energy density
 to the dark matter energy density,
 which is based on a unified dark matter model
 \cite{Bertacca:2010mt,Frion:2023xwq}.
The fluid model has a certain level of consistency:
 the weak energy condition $\rho+p>0$ is satisfied, for example.
As it is described in \cite{Bertacca:2010mt},
 this fluid model should be represented as a scalar field theory
 with some non-trivial kinetic term and potential,
 namely this model should be a sort of quintessence model.

The investigation in this article is still at the level of background evolution.
It is very interesting to investigate perturbations in this model:
 the CMB perturbations, the matter density perturbations
 (and also various physics in galaxy survey) should be especially interesting.
These investigation is even necessary to judge
 the viability of this idea as a solution of the problem of Hubble tension.
We leave this investigation
 requiring an extensive change in dark sector from that of the $\Lambda$CDM model
 for future works.
The redshift value of transition
 should affect the star and galaxy formation,
 which could relate with the recent JWST observations of the Balmer break galaxies
 at very high redshifts
 \cite{Labbe:2022ahb,Boylan-Kolchin:2022kae,Forconi:2023izg,
       Xiao:2023xxx,Desprez:2023pif,Vikaeus:2023cyi}.
It would be interesting,
 if this non-standard dark sector
 could give a hint for the formation of supermassive blackholes. 

In the standard consideration,
 the dark matter is introduced as some particle
 to explain CMB acoustic oscillation and later structure formation,
 and the dark energy is introduced
 to describe late-time accelerating expansion of the universe.
In this work
 we have escaped from this standard picture
 and examine the dark sector without such prejudice.
Originally this is the concept of the unified dark energy models
 \cite{Bertacca:2010mt,Frion:2023xwq}.
We await the near future precise observational information,
 especially for the values of $\Omega_m$ as well as $H_0$,
 keeping in mind that the experimental value could change in time
 as shown in History Plot in \cite{ParticleDataGroup:2020ssz}.

\section*{Acknowledgments}

This work was supported in part by JSPS KAKENHI Grant Number 19K03851.

\end{document}